\documentclass[11pt,twoside]{article}

\usepackage{asp2006}
\usepackage{epsf}
\usepackage{psfig}
\usepackage{lscape}

\newcommand{\arcs}{\hbox{$^{\prime\prime}$}}

\newcommand{\chandra}{\emph{Chandra}}
\newcommand{\xmm}{\emph{XMM-Newton}}
\newcommand{\Hi}{H\textsc{i}}
\newcommand{\Ha}{\ensuremath{\mathrm{H}_{\alpha}}}
\newcommand{\s}{\ensuremath{\mbox{~s}}}
\newcommand{\ps}{\ensuremath{\s^{-1}}}
\newcommand{\km}{\ensuremath{\mbox{~km}}}
\newcommand{\kmps}{\ensuremath{\km \ps}}

\newcommand{\Msol}{\ensuremath{M_{\odot}}}

\begin{document}
\title{A joint GMRT/X-ray study of galaxy groups}   
\author{E. O'Sullivan$^1$, S. Giacintucci$^1$, J.~M. Vrtilek$^1$, S. Raychaudhury$^2$, R. Athreya$^3$, T. Venturi$^4$ and L.~P. David$^1$} 
\affil{$^1$Harvard-Smithsonian Center for Astrophysics, USA \\
$^2$University of Birmingham, UK \\
$^3$NCRA--TIFR Pune, India  \\  
$^4$INAF--IRA Bologna, Italy} 

\begin{abstract} 
  We present results from combined low--frequency radio and X--ray studies
  of nearby galaxy groups. We consider two main areas: firstly, the
  evolutionary process from spiral--dominated, \Hi--rich groups to
  elliptical--dominated systems with hot, X--ray emitting gas halos;
  secondly, the mechanism of AGN feedback which appears to balance
  radiative cooling of the hot halos of evolved groups.  The combination of
  radio and X--ray observations provides a powerful tool for these studies,
  allowing examination of gas in both hot and cool phases, and of the
  effects of shock heating and AGN outbursts.  Low-frequency radio data are
  effective in detecting older and less energetic electron populations and
  are therefore vital for the determination of the energetics and history
  of such events.  We present results from our ongoing study of Stephan's
  Quintet, a spiral--rich group in which tidal interactions and shock
  heating appear to be transforming \Hi\ in the galaxies into a diffuse
  X--ray emitting halo, and show examples of AGN feedback from our sample
  of elliptical--dominated groups, where multi--band low--frequency radio
  data have proved particularly useful.
\end{abstract}


\section{Introduction}   
Feedback processes play an important role in governing the development and
evolution of galaxies and the groups and clusters in which they reside. In
galaxy clusters AGN-driven radio sources appear to be the primary means of
balancing cooling in the hot X-ray emitting gas halo, and thus preventing
high rates of star formation which are not observed. However, the impact of
feedback processes may be most important in less massive galaxy groups,
where the majority of galaxies -- and the majority of baryonic matter in
the Universe -- reside \citep{Ekeetal04}. To understand the
mechanisms by which feedback operates in groups and the effect of the
AGN/hot gas interactions, we have begun a study of a number of nearby
systems using high--quality multi--frequency GMRT observations and deep
archival Chandra and XMM-Newton X-ray data. Of particular interest are: i)
the relationship between the observed X--ray and radio structures, ii) the
mechanisms of energy injection, and iii) the properties of the radio
galaxies, including their ages and duty cycles.

Galaxy groups are a diverse class of systems, ranging from loosely bound,
recently virialised systems to compact formations in which galaxy
interactions are common. A broad distinction can be drawn between
spiral--rich and elliptical--dominated systems: the latter commonly possess
an extended halo of hot, X--ray emitting gas, little \Hi\ except in
galaxies at large radii, and typically have a central FR-I (or more rarely
FR-II) radio galaxy. Spiral--rich groups are generally X--ray faint, but
contain much larger amounts of \Hi\ in the galaxies or in tidal features.
The merging process by which spirals in these systems are transformed into
ellipticals is relatively well understood, but it is as yet unclear what
happens to the cold gas in the galaxies, or how the hot halo is formed. To
investigate these issues we have selected a small sample of spiral--rich
groups currently undergoing rapid evolution. We discuss results from the
first of these below.

\section{Stephan's Quintet}
\begin{figure}[!t]
\centerline{
\psfig{file=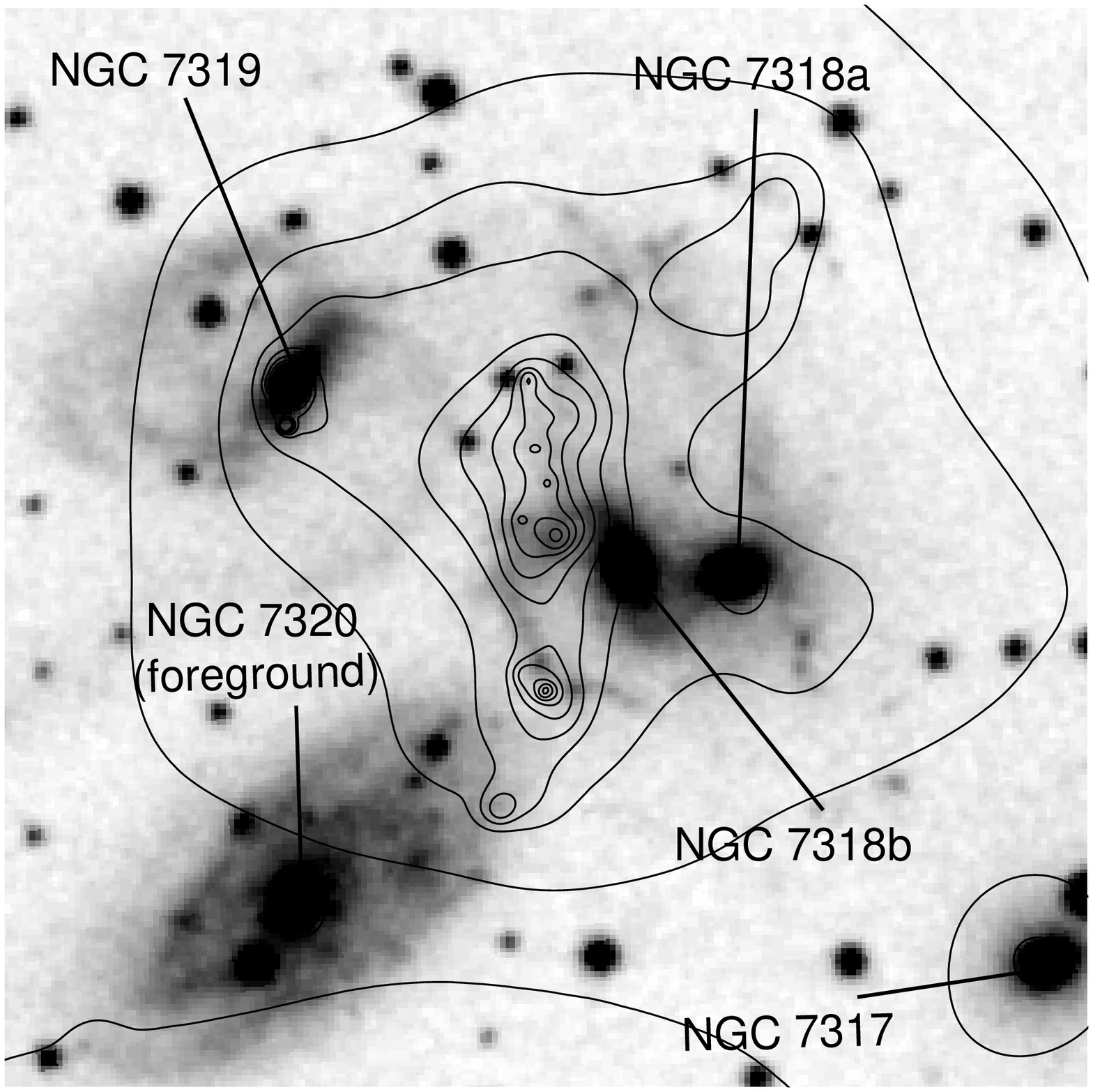,width=5cm}
\hspace{5mm}
\psfig{file=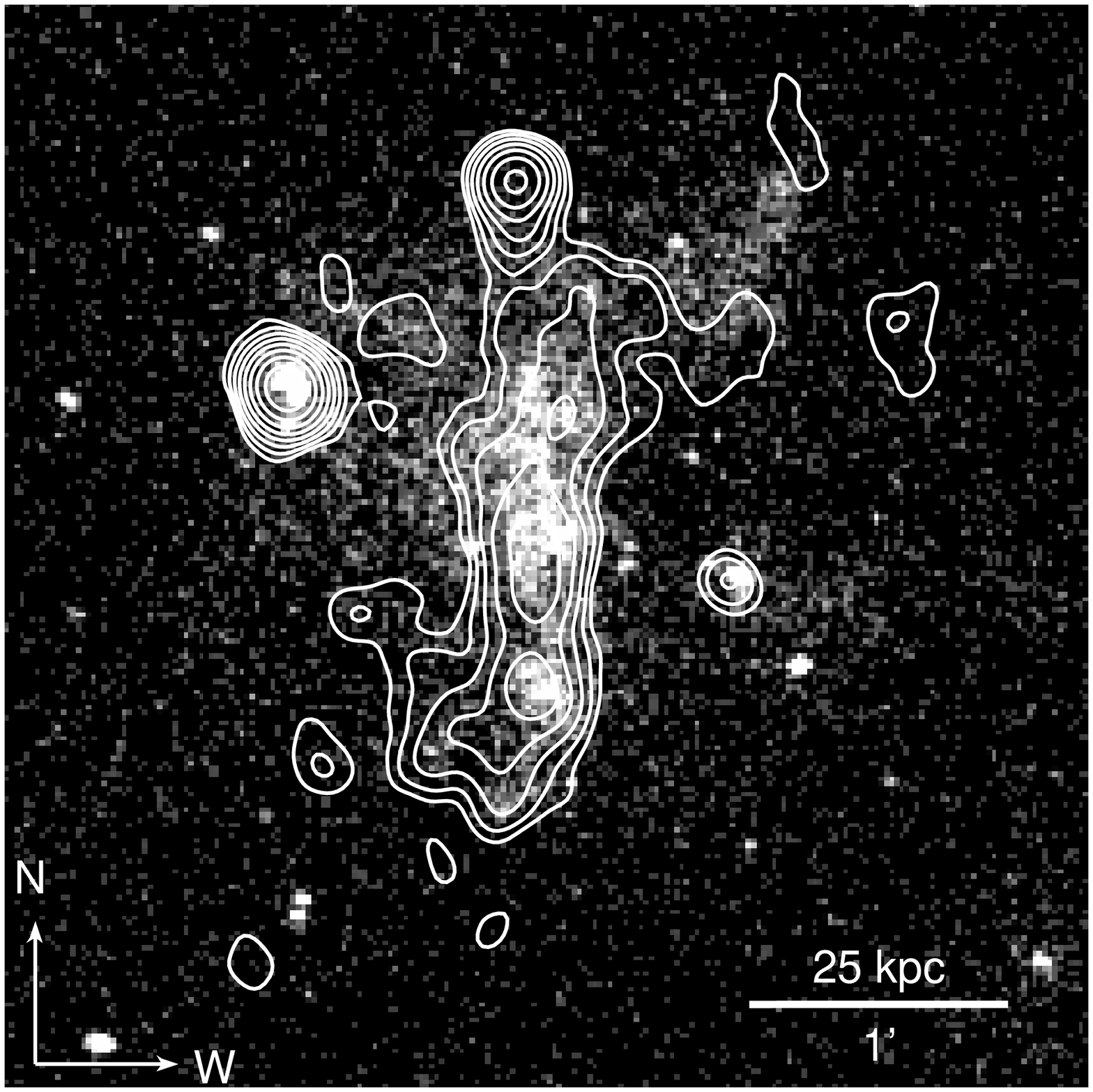,width=5cm}
}
\caption{\label{ejosfig1} \textit{Left:} Smoothed \textit{Chandra} X-ray
  contours overlaid on POSS2 $B_j$-band image of Stephan's Quintet, with
  principal galaxies marked.  \textit{Right:} GMRT 610~MHz contours
  overlaid on \textit{Chandra} 0.3-2.0~keV image of the group. Contours are
  spaced by factors of 2, beginning at 3$\times$r.m.s.=0.36 mJy/b. The HPBW
  is 6$^{\prime\prime}\times$5$^{\prime\prime}$. The panels have identical
  scale and alignment.  }
\end{figure}

Stephan's Quintet (HCG~92) is a galaxy group consisting of five main
members, of which four are located in a compact core, with one foreground
spiral galaxy superimposed (see Fig.~\ref{ejosfig1}). The group is a rare
example of an ongoing high-velocity galaxy interaction; NGC~7318b, a
near-edge-on spiral galaxy, is passing through the group core with a
velocity of $\sim$850\kmps. A ridge of radio continuum emission is observed
at 1.4~GHz, coincident with the eastern edge of the galaxy, and this has
been interpreted as a shock caused by the collision of NGC~7318b with
tidally stripped \Hi\ \citep[see][for a summary]{Sulenticetal01}. This
feature is observed in several wavebands, most notably in \Ha\ and X--ray
\citep[e.g.,][]{Trinchierietal05}. The group also shows evidence of past
interactions, including stellar and \Hi\ tidal tails, thought to have been
formed by one or more passages through the group core by a small spiral
galaxy, NGC~7320c \citep[][not shown in Fig.~\ref{ejosfig1}]{Molesetal97}.
The \Hi\ with which NGC~7318b is colliding is likely material stripped from
the galaxies during these interactions.

\begin{figure}
\centerline{ 
\psfig{file=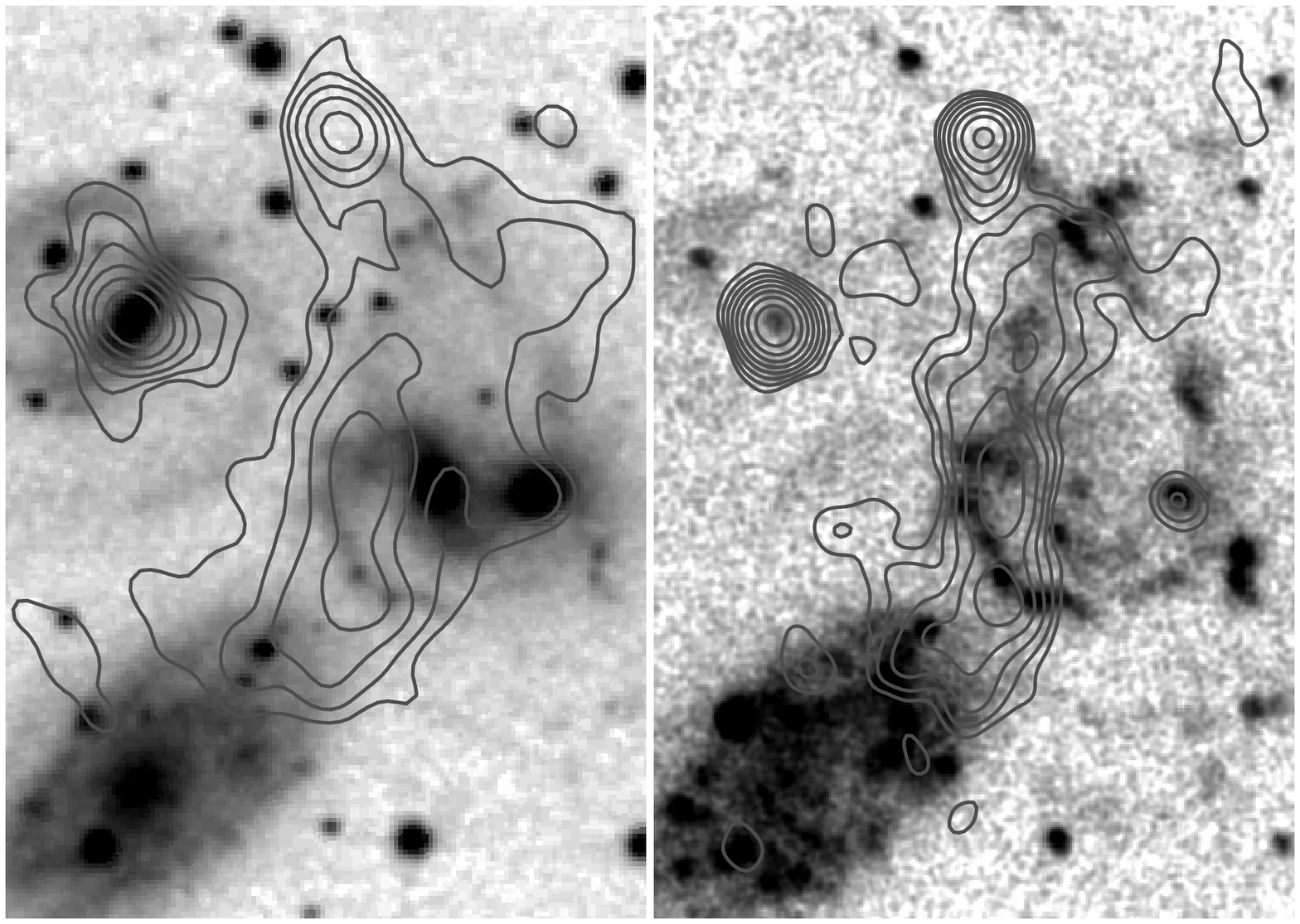,width=7cm}
\frame{\psfig{file=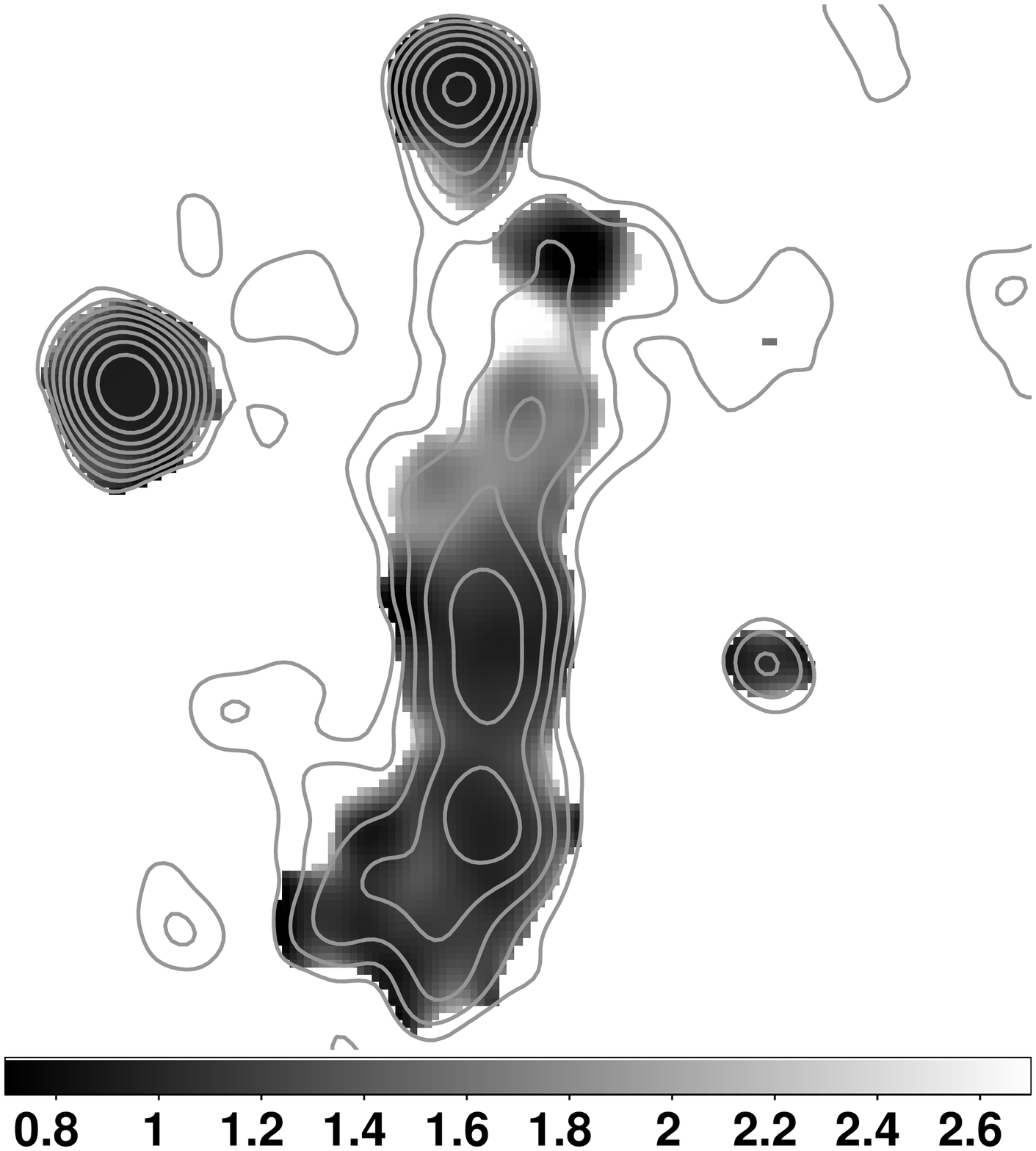,width=4.3225cm}}
}
\caption{\label{ejosfig2} {\itshape Left:\/} 327~MHz GMRT contours overlaid
  on POSS blue image of the core of Stephan's Quintet. Contours start from
  3$\times$r.m.s.=0.9mJy/b, HPBW is
  12$^{\prime\prime}\times$10$^{\prime\prime}$.  {\itshape Center:\/}
  610~MHz GMRT contours overlaid on a \textit{Swift} UVOT UVW2-band image
  ($\sim$200~nm).  {\itshape Right:\/} 1400-610~MHz spectral index map
  (HPBW of 6$^{\prime\prime}\times$5$^{\prime\prime}$) with 610~MHz
  contours overlaid.}
\end{figure}

We have examined the structure of the ridge and the surrounding diffuse gas
using a combination of GMRT 610 and 327~MHz, and VLA 1.4~GHz radio
observations and a deep ($\sim$93~ks) \chandra\ pointing
\citep{OSullivanetal08}.  Fig.~\ref{ejosfig1} shows the location and
structure of the ridge, and a comparison of the relative structures in
radio and X--ray bands. The X--ray emission is brightest in the northern
part of the ridge, while the radio emission is brightest in the south.
Fig.~\ref{ejosfig2} shows GMRT 327 and 610~MHz contours overlaid on optical
and UV images of the group core, and on a 1400-610~MHz spectral index map.
The diffuse radio emission is more extended at lower frequencies,
including areas west of the ridge and emphasizing the brightness of the
southern part of the ridge. The spectral index map shows similar structure;
the southern part of the ridge has an index of 0.7-0.8, while the northern
part has a steeper spectrum.  The brightest radio emission corresponds with
knots of UV emission in the south-eastern spiral arm of NGC~7318b and as
the spectral index in this region is consistent with star formation it
seems likely that the underlying extended radio emission arising from
shocks is enhanced by star formation in some areas.

From the large-scale diffuse X--ray emission, we can estimate the extent of
the gas halo ($\sim$80~kpc) and its total mass,
$\sim$2.8$\times10^{10}$\Msol. This is similar to the estimated \Hi\
deficit of the group
\citep[$\sim$2$\times10^{10}$\Msol,][]{Verdes-montenegroetal01}, suggesting
that the majority of the hot gas in the system may have been formed by
shock heating \Hi. However, the time taken for such gas to expand at the
sound speed from the region of the ridge ($\sim$125~Myr) is much longer
than the total time for NGC~7318b to pass through the group
\citep[20-80~Myr][]{Sulenticetal01}. It therefore seems likely that
multiple episodes of shock heating are necessary. The prior interactions
with NGC~7320c could be responsible.

\section{AGN feedback in elliptical-dominated groups}
\begin{figure}[!t]
\centerline{
\psfig{file=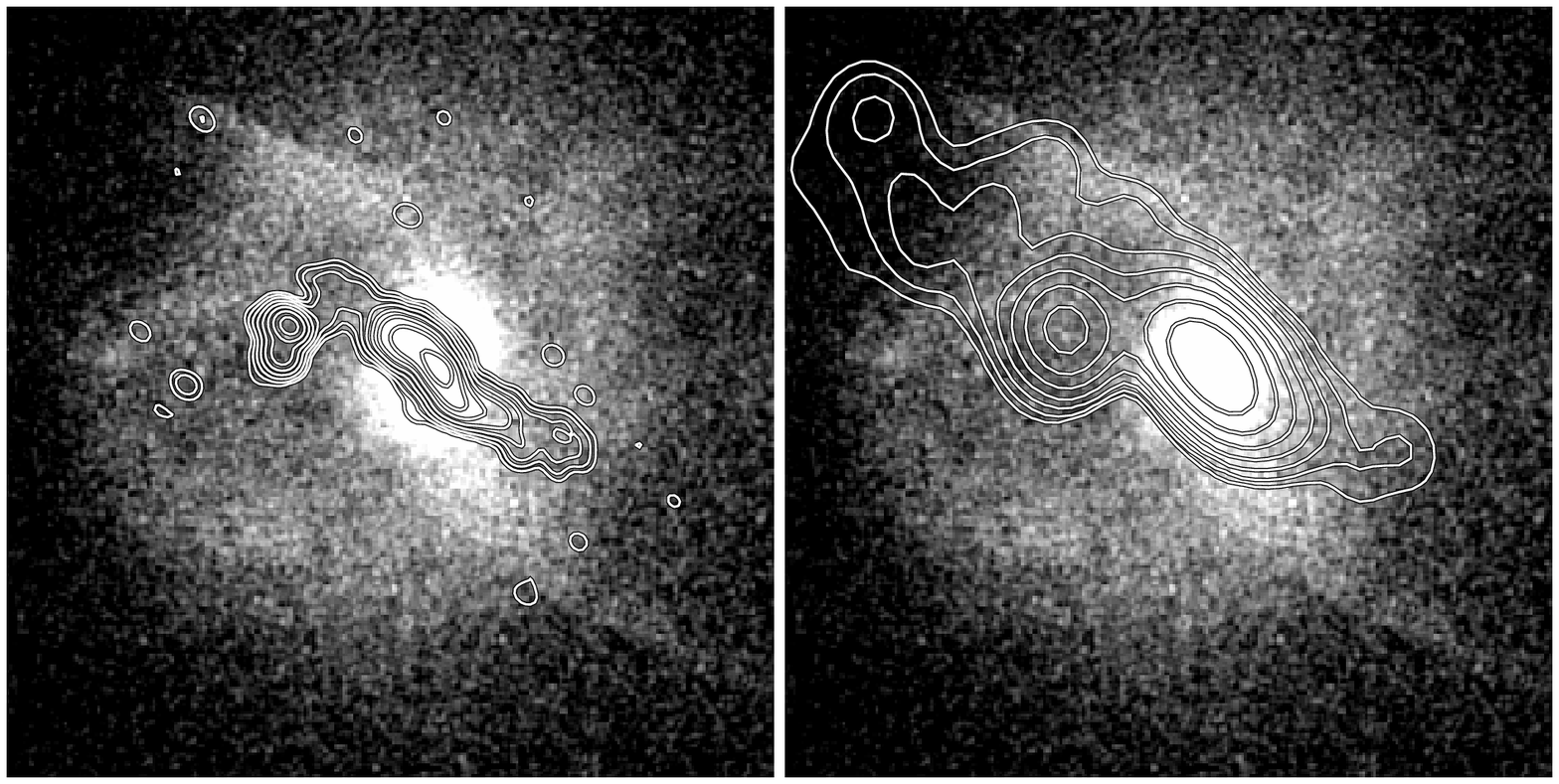,height=4.45cm}
\psfig{file=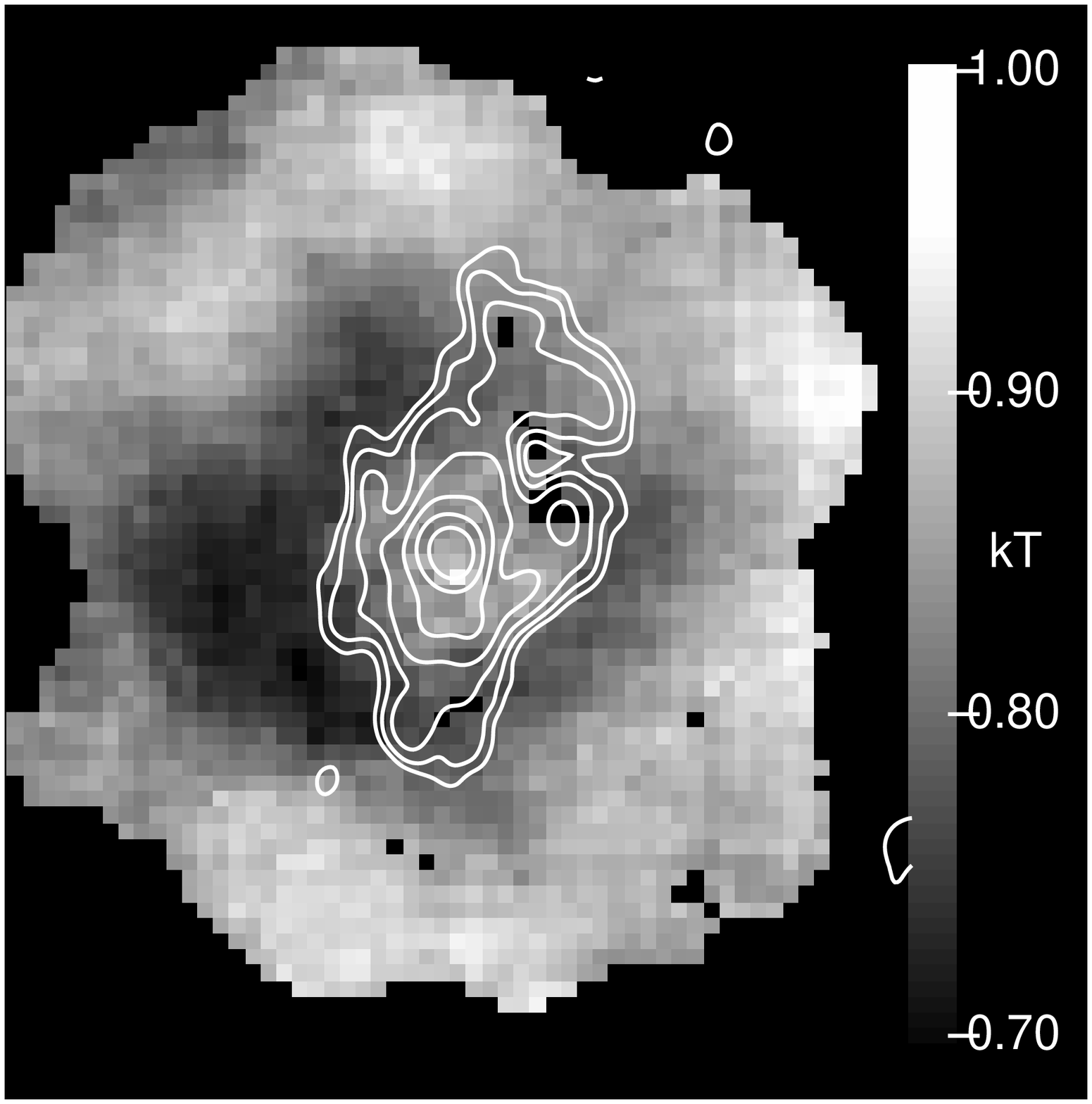,height=4.45cm}
}
\caption{\label{ejosfig3} \textit{Left:} GMRT 610~MHz contours overlaid on
  \textit{Chandra} 0.5-3.0~keV image of NGC~4636. Contours start at
  3$\times$r.m.s.=0.15~mJy/b, HPBW is 6\arcs$\times$4\arcs.
  \textit{Centre:} GMRT 235~MHz contours on the \textit{Chandra} image,
  3$\times$r.m.s.=0.6~mJy/b, HPBW 16\arcs$\times$13\arcs.  \textit{Right:}
  \xmm\ X-ray temperature map of the core of the NGC~3411 group (in units
  of keV) with GMRT 610~MHz contours overlaid. 3$\times$r.m.s.=0.39~mJy/b,
  HPBW 18\arcs$\times$15\arcs.}
\end{figure}

we have also assembled a sample of 18 X--ray bright elliptical--dominated
groups in order to examine the mechanisms and effects of AGN feedback in
such systems. While this is still a work in progress, several systems
demonstrate the benefits of our combined approach. In NGC~4636, \chandra\
and \xmm\ observations have revealed a complex of shocks and cavities in
the galaxy core \citep[][and references therein]{OSullivanetal05}, but VLA
1.4~GHz observations detect only small-scale jets \citep{Allenetal06}. Our
GMRT 610 and 235~MHz observations (Fig.~\ref{ejosfig3}) reveal more
extended emission which correlates well with the X-ray structure. This
suggests that the AGN of NGC~4636 is restarting, with young jets pushing
out into old lobes/cavities from a previous outburst.

NGC~507 provides an example of an old source where no new activity has yet
begun. No evidence of jets is found at any frequency, suggesting that the
AGN is currently inactive, with the lobes passively aging. Modelling of the
radio spectrum \citep{Murgiaetal09} supports this picture and indicates
that the source has likely been in this `dying' phase for $\sim$30\% of its
total lifespan. However, X--ray/radio interactions are still playing an
important role; without the surrounding hot IGM, the lobes would likely
have expanded and diffused, reducing their surface brightness and
preventing their detection.

A few systems in the sample show examples of interactions which do not
follow the common jet/cavity mechanism. In the NGC~3411 group, X--ray
temperature maps reveal a hot core region surrounded by a cool shell of
gas, strongly suggesting that an AGN outburst has heated the inner group
halo \citep{OSullivanetal07}.  VLA 1.4 and 5~GHz maps show no evidence of
jets and very little extended emission, but GMRT observations reveal a much
larger scale structure which correlates closely with the heated region (see
Fig.~\ref{ejosfig3}). The lack of clearly defined lobes and jets could be
explained if the lobes were aligned along the line of sight, but this would
produce a central X--ray surface brightness deficit, which is not observed.
X--ray spectral fitting finds no evidence for a significant
inverse--Compton component. It therefore seems likely that AGN heating has
occurred through a different mechanism, perhaps involving the disruption of
the jets on very small scales, leading to mixing of radio and X--ray plasma
in the group core.  Further observations are required to test this
hypothesis, but if accurate, this would provide a new means to reheat
cooling gas in the centres of groups without causing strong disturbances.

\section{Conclusions}
The results presented here demonstrate the power of combining
low--frequency radio and X--ray observations to examine AGN
feedback, shock heating and star--formation in galaxy groups. In Stephan's
Quintet we may have an example of the type of interactions through which
the hot halos of elliptical--dominated groups are built up. The
correspondence between the mass of hot gas in the system and the \Hi\
deficit suggests that the hot halos of groups may initially form from cool
material. The extensive shock heating caused by the collision between
NGC~7318b and the \Hi\ filament provides an clear mechanism by which this
may happen; tidal interactions strip \Hi\ from the galaxies and the
resulting intergalactic clouds are heated by shocks associated with galaxy
merger or capture events. Studies of similar systems should help determine
the importance of this mechanism to group evolution.

Examples from our sample of 18 elliptical--dominated groups show the
benefits of extending observations to lower frequencies. As well as
allowing us to detect faint extended radio structures which go unseen at
higher frequencies, radio spectral fitting provides estimates of the age of
sources and the state of the plasma in various regions; `dying' galaxies in
which nuclear activity has ceased can be identified, while in restarting
sources remnant emission from prior outbursts can be separated from those
regions powered by the new jets. This helps us to understand both the
mechanisms of AGN feedback and the energies and timescales involved. Such
information is key to determining the thermal history of individual
systems, and of fundamental importance to our understanding of the role of
AGN in galaxy and group evolution.

\acknowledgements 
The authors thank D.~J. Saikia and S. Immler for
providing access to the GMRT and Swift data for Stephan's Quintet. Support
for this work was provided by NASA through Chandra Award G07-8133X-R.


\end{document}